\documentclass[11pt,a4paper]{article} 

\usepackage[T1]{fontenc}
\usepackage[utf8]{inputenc}   
\usepackage{lmodern}          
\usepackage{microtype}        
\usepackage{amsmath,amssymb,mathtools}
\usepackage{graphicx}
\usepackage{siunitx}          
\usepackage{authblk}          
\usepackage[colorlinks=true,linkcolor=blue,citecolor=blue,urlcolor=blue]{hyperref}
\usepackage[margin=2.5cm]{geometry} 
\usepackage{placeins}
\usepackage{caption}
\usepackage{subcaption}

\title{Late-Time Power Laws in Fluorescence Decays:\\ Quantum Decay or Alternative Mechanisms?}

\author[1,2]{Francesco Giacosa}
\author[3]{Anna Kolbus}
\author[1]{Krzysztof Kyziol}
\author[4]{Magdalena Plodowska} 
\author[1]{Milena Piotrowska\thanks{corresponding author: milena.piotrowska@ujk.edu.pl}}
\author[1]{Karol Szary}
\author[1]{Arthur Vereijken}

\affil[1]{Jan Kochanowski University, Institute of Physics, Faculty of Natural Sciences, ul. Uniwersytecka 7, 25-406 Kielce, Poland}
\affil[2]{J.W. Goethe University, Institute of Theoretical Physics, Department of Physics, Max-von-Laue Str. 1, 60438 Frankfurt am Main, Germany}
\affil[3]{Jan Kochanowski University, Institute of Chemistry, Faculty of Natural Sciences, ul. Uniwersytecka 7, 25-406 Kielce, Poland}
\affil[4]{Jan Kochanowski University, Institute of Biology, Faculty of Natural Sciences, ul. Uniwersytecka 7, 25-406 Kielce, Poland}

\begin{document}

\maketitle

\begin{abstract}
    In quantum mechanics, the properties of an unstable system are closely related to its corresponding energy distribution. One of the predictions of the theory is the emergence of a~late-time power law in the survival probability $P(t)$ of the unstable state. However, in complex systems, similar effects may emerge due to other phenomena. This is the case for nonextensive statistical models, for some continuous lifetime distribution models, and also for certain energy-migration models applied to fluorescence. In this work, these alternative interpretations are confronted with data from time-resolved fluorescence spectroscopy experiments with erythrosine B and eosine Y as fluorophores. It turns out that some of the features expected within the quantum-mechanical description, such as the sharp transition from the exponential to the power-law regime, cannot be reproduced by the specific statistical and energy-migration parametrizations fitted in the analysis.
\end{abstract}


\date{ }

\section{Introduction}

One of the most established laws in Physics is the exponential decay law. It is derived at a~phenomenological level under the assumption that past events cannot influence the decay rate. Therefore, no memory effects are present and, as a result, decay processes closely resemble Markovian processes. The exponential law follows when the decay probability per unit time, conditional on survival, is constant. However, the quantum mechanical (QM) treatment of decays predicts small but qualitatively important different results. For a time-independent Hamiltonian, e.g. for a Lee-Friedrichs Hamiltonian \cite{Lee:1954iq,friedrichs1948}, the survival amplitude is closely related to the energy distribution $\rho(E)$ of an unstable system \cite{ghirardi,dicus2002}:
\begin{equation}
    A(t) = \int \mathrm{d}E \; \rho (E) \exp \left( -i \frac{E}{\hbar} t \right) \; ,
\end{equation}
with the survival probability $P(t)$ given by the modulus squared of the amplitude:
\begin{equation}
    P(t) = \left| A(t) \right| ^2 \; .
\end{equation}
The fluorescence intensity $I(t)$, discussed extensively in the next sections, is proportional to $-P'(t)$, with the initial number of excited states $N_0$ as an appropriate proportionality factor:
\begin{equation}
    I(t) = -N_0 \frac{\mathrm{d}P}{\mathrm{d}t} \; .
\end{equation}
Deviations from the exponential decay law are expected both at very short and very long times; see, e.g., Ref. \cite{ghirardi} for a theoretical overview, Refs. \cite{khalfin,Winter:1961zz,Peshkin:2014jdw,dicus2002,delaMadrid:2015ooa,Giacosa:2019jxz} for specific QM results based on tunnelling with time-independent potentials, and Refs. \cite{Facchi:1999ik,Giacosa:2010br,Giacosa:2011xa,Giacosa:2021hgl} for quantum field theoretical (QFT) approaches. \smallskip \newline

Experimental verification of deviations at short times was achieved in engineered systems in Refs. \cite{Wilkinson:1997sez,Crespi}, while long-time deviations were observed using fluorescent molecules in Refs. \cite{Rothe:2006rma,RID_paper}; see below. Indeed, for elementary systems these deviations are typically too small to be detected, as illustrated by the case of the $2P$--$1S$ transition of the H atom \cite{Facchi:1998abc,Giacosa:2024yhn}. For hidden evidence of nonexponential behaviour in beryllium nuclear decays, see Ref. \cite{Kelkar:2004zz}. \smallskip \newline

Short-time deviations may also lead to the quantum Zeno effect (QZE) \cite{Misra:1976by,Facchi:2008nrb}: for a finite energy variance $\sigma_E^2$, one has $P'(t)|_{t=0}=0$, and sufficiently frequent measurements may slow down the decay by forcing the system to remain in its initial state \cite{Itano:1990zz,ZenoUnstable}. \smallskip \newline

On the other hand, the existence of a lower bound in the energy spectrum, i.e. a minimal admissible energy of the system, leads to a power-law decay at long times,
$P(t) \sim t^{-\gamma}$, where $\gamma$ depends on the behaviour of the spectral function $\rho(E)$ near the threshold (for details, see Ref. \cite{PhysRevA.77.042719}). This effect is well understood theoretically, as it emerges from fundamental properties of Fourier transforms \cite{khalfin}. However, experimental evidence in direct temporal measurements has so far been observed only in complex fluorescent substances \cite{Rothe:2006rma,RID_paper}, and not for all tested substances \cite{Giacosa:2025AO}. In Ref. \cite{RID_paper}, there are indications of a quite intriguing quantum phenomenon: the power-law exponent depends on the channel in which it is observed. This result is in agreement with the theory of the multichannel decay law developed in Ref. \cite{Giacosa:2021hgl} and initiated in Ref. \cite{Giacosa:2011xa}, which applies to both QM and QFT frameworks. For an ideal single-channel radiative decay, the emitted photon rate is proportional to $-P'(t)$. In the multichannel case, $-P'(t)$ gives the total decay rate, whereas a selected detection band probes the corresponding partial radiative rates \cite{Giacosa:2021hgl}. \smallskip \newline

However, the fact that experiments aimed at observing the late-time power-law involve complex systems (and are, up to now, not practically feasible using simple elementary systems, such as atomic states or elementary particles) means that an explanation in terms of fundamental QM effects is not the only possibility. The emergence of a power-law may also be caused by different lifetime distributions, non-extensive statistics, or energy-migration processes. In this work, these possibilities are compared with experimental data obtained from time-correlated single-photon counting (TCSPC) spectroscopy measurements.

\section{Testing alternative origins of the late-time power law}
In this Section, after a brief description of the experiment, we discuss alternative statistical and photophysical mechanisms that could generate a power law at late times. 

\subsection{Brief description of the experimental setup}
The experimental setup consists of a few individual components. First, a pulsed laser source is employed. In our case, this is realized by the PicoQuant Laser Combining Unit (LCU) with two individual picosecond laser diodes -- PicoQuant LDH-D-C-440 (wavelength of 438 nm) and PicoQuant LDH-D-C-485 (wavelength of 483 nm). The pulse widths of both diodes are less than 120 ps, while spectral widths range from 2 to 8 nm. In the experiment, only the PicoQuant LDH-D-C-440 was used. The inverted confocal microscope, Nikon Eclipse Ti-E, serves as an optical component. It is responsible for the focusing of the laser beam, and also for the formation and guidance of the beam emitted from a sample towards a detection unit. The detection setup consists of two identical PicoQuant PMA Hybrid 40 hybrid detectors separated by a dichroic mirror. Each of them is equipped with an individual bandpass filter (520/35 and 600/50 nm, respectively). The last component is the PicoHarp 300 module, which is the heart of the TCSPC spectroscopy. It performs the determination of the time for each individual detection event. The assignment of each photon to discrete time bins results in the formation of the histogram corresponding to the fluorescence decay curve. \smallskip \newline

As described in \cite{RID_paper}, both erythrosine B and eosine Y dissolved in methanol and water with a concentration of $10^{-5}$ mol/dm$^3$ show a late-time power-law decay. Moreover, the power-law exponent is channel-dependent; it was shown that this behavior is in agreement with QM. However, a discussion of other potential mechanisms that may explain the formation of similar late-time structures is essential for the proper interpretation of the experimental results. The next sections of this article are devoted to this question. \smallskip \newline
During the experiment, the laser operates with the frequency of 10 MHz leading to the period of 100 ns. The time interval is then divided into 3125 bins, each of width 0.032 ns. For each individual measurement, the total acquisition time is equal to 10 minutes (or 15 minutes in some particular cases). Datasets from distinct trials are finally combined into one table in a way that takes into account the geometrical properties of the setup that may differ between measurements.

\subsection{Continuous lifetime distribution}

In TCSPC spectroscopy, photon arrival times are recorded with respect to the excitation pulse and accumulated over many laser cycles, thereby forming the fluorescence decay curve. In order to avoid pile-up effects, the detection probability is kept sufficiently low so that typically no more than one photon is registered per excitation cycle \cite{lakowicz}. In complex molecular systems, however, the fluorescence decay does not necessarily have to be characterised by a single lifetime. Fluctuations in the molecular configuration, caused for instance by thermal excitations, as well as variations in the local solvent environment, may lead to a distribution of effective lifetimes. Under such conditions, the standard multiexponential description may be replaced by a continuous lifetime distribution as an effective model. This approach is based on the assumption that for each individual decay event the effective decay width is basically constant. The fluorescence decay function can then be written in the form
\begin{equation}
    I(t) = I_0 \int_{0}^{\infty} \mathrm{d}\tau \,
    \rho(\tau)\exp\left(-\frac{t}{\tau}\right) \; ,
\end{equation}
where $\rho(\tau)$ denotes the distribution of fluorescence lifetimes. This equation provides, in principle, a rather general framework that can be directly confronted with experimental data once a specific form of $\rho(\tau)$ is assumed. Common choices include, for instance, Gaussian (restricted to positive lifetimes) or lognormal distributions. Although such distributions do not generally reproduce an exact power law at asymptotically long times, sufficiently broad lifetime distributions, possibly together with the experimental background, may mimic power-law-like behaviour over a finite time interval. \smallskip \newline

A general determination of $\rho(\tau)$ from the measured fluorescence decay curve is, however, an ill-conditioned inverse problem due to the functional freedom. For this reason, rather than attempting a model-independent reconstruction of $\rho(\tau)$, in the following we focus on simple analytical models with a small number of parameters. Some of these models are closely related to continuous lifetime distributions and explicitly generate power-law behaviour at late or intermediate times.

\subsection{Non-extensive Tsallis statistics}

One of the most interesting effects that may lead to power-law decays originates from non-extensive Tsallis statistics \cite{Tsallis:1987eu}. It provides an effective description of non-equilibrium systems and may be appropriate for strongly heterogeneous media with long-range correlations. The theory is based on an alternative, non-additive definition of entropy:
\begin{equation}
    S_q = k_B \frac{1-\sum_{n=1}^{w} p_n^q}{q-1} \; , 
    \qquad q \in \mathbb{R} \; ,
\end{equation}
with $p_n$ as the probability that the $n$-th molecular state is occupied and $w$ as the total number of available states. The limit $q \to 1$ corresponds to the standard Boltzmann--Gibbs entropy. A~commonly used relaxation function associated with this framework is the q-exponential, which we adopt here as an effective parametrization of the fluorescence decay:
\begin{equation}
    \label{eq:Tsallis_fluorescence-intensity}
    I(t) = I_0 \left[ 1-(1-q)\frac{t}{\tau_q} \right]^{\frac{1}{1-q}} \; ,
\end{equation}
where the parameter $\tau_q>0$ replaces the classical exponential decay lifetime. For $q>1$, this formalism reproduces a power-law tail at late times:
\begin{equation}
    I(t) \sim  t^{-\frac{1}{q-1}} \; .
\end{equation}
Such an effective description may also arise from fluctuations of the decay rate following a gamma distribution \cite{wilkj2000}, with the parameter $q$ providing a measure of the relative strength of these fluctuations.
We test the model below, but in order to reliably reproduce experimental conditions we introduce the constant background level $B$ and the parameter $t_0$ fixed at the maximum of the intensity since the reference point of timescale is rather due to laser pulse than signal itself:
\begin{equation}
    \label{eq:fitting_function}
    I_{\mathrm{fit}}(t) = I_{\mathrm{model}}(t-t_0) + B
\end{equation}
The quantity $I_{\mathrm{model}}(t)$ corresponds directly to Eq. (\ref{eq:Tsallis_fluorescence-intensity}). The relation between the model and the fitting functions, expressed in Eq. (\ref{eq:fitting_function}), preserves its form for all parametrizations considered in this article. The same applies to error estimations. Assuming that the number of counts in each bin follows the Poisson statistics, the $\chi^2$ value is evaluated using the weighted least squares statistic of the following form:
\begin{equation}
    \chi ^2 = \sum _ i \frac{\left[ I_{\mathrm{fit} (t_i)} - I_i \right] ^2}{I_i} \; ,
\end{equation}
where $I_i$ denotes the number of counts measured in a bin corresponding to the time $t_i$. Within the quadratic approximation around the minimum,
the parameter covariance matrix is $\Sigma=2H^{-1}$, where the Hessian is evaluated at the best-fit parameters. The corresponding standard uncertainties are $\sigma_{p_i}=\sqrt{\Sigma_{ii}}$.
\smallskip \newline

\begin{figure}[!htbp]
    \centering
    \begin{subfigure}{0.48\textwidth}
        \captionsetup{width=0.8\textwidth}
        \includegraphics[width=1\linewidth]{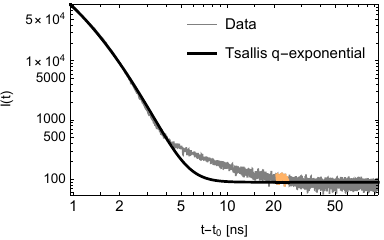}
        \caption{Erythrosine B, channel 1 -- $\chi_{\nu}^2 = 6.79$, $q = 1.132 \pm 0.001$, $\tau_q = 0.262 \pm 0.002 \, \mathrm{ns}$.}
        \label{fig:Tsallis-erythrosine}
    \end{subfigure}
    \begin{subfigure}{0.48\textwidth}
        \captionsetup{width=0.8\textwidth}
        \includegraphics[width=1\linewidth]{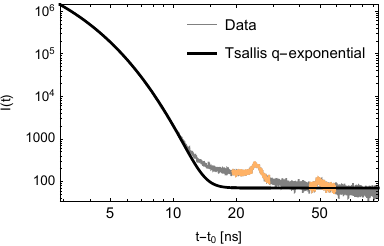}
        \caption{Eosine Y, channel 1 -- $\chi_{\nu}^2 = 9.80$, $q = 1.0135 \pm 0.0002$, $\tau_q = 1.0213 \pm 0.0005 \, \mathrm{ns}$.}
        \label{fig:Tsallis-eosine}
    \end{subfigure}
    \caption{Results of the fits using the Tsallis-statistics model. Non-fitted data are marked in orange, see Ref. \cite{RID_paper} -- this is due to effects related to time-to-digital converter and its internal reference clock leading to periodic structures with the period of 25 ns.}
    \label{fig:Tsallis}
\end{figure}

The results of the analysis show that this model does not reproduce the shape of the experimental fluorescence decay curves. In particular, the change in slope around the turnover from the exponential to the power-law regime is not sufficiently pronounced. The results for erythrosine B and eosine Y, considering only channel 1, are presented in Fig.~\ref{fig:Tsallis}.

\subsection{Delayed fluorescence}

Delayed fluorescence is a particularly interesting process that can indeed lead to power-law behaviour of the fluorescence intensity decay curve. Such effects may arise from nonlinear kinetics of triplet states caused by second-order triplet relaxation processes \cite{Croizat2020}. In the first-order mechanism considered here, molecules found initially in a triplet state are thermally excited to a higher singlet state, followed by radiative relaxation to $S_0$,
$T_1 \xrightarrow{k_BT} S_1 \xrightarrow{h\nu} S_0$.
On the other hand, second-order processes are associated with a different mechanism. They are caused by collisions of two molecules in the triplet state, resulting in one molecule in the ground state and the other in an excited singlet state,
$2T_1 \xrightarrow{} S_0+S_1$.
For this reason, the contribution of these processes scales as the square of the triplet-state concentration $[T_1]^2$, and the argument that delayed fluorescence occurs on much longer timescales does not necessarily apply here. Unlike a~purely first-order process, the characteristic timescale of triplet-triplet annihilation depends on the triplet concentration. The solution of the differential equation governing the kinetics,
\begin{equation}
    \frac{\mathrm{d}[T_1]}{\mathrm{d}t} = -b[T_1]^2 \; ,
\end{equation}
is given by \cite{Hayer}
\begin{equation}
    [T_1](t) = \frac{[T_1]_0}{1+b[T_1]_0 t} \; .
\end{equation}
Thus, at large times, the triplet population follows a power law, $[T_1](t)\sim t^{-1}$, and its relevance to the observed tails must be assessed through the product $b[T_1]_0$. If the regenerated singlet state relaxes rapidly compared with the evolution of the triplet population, the delayed-fluorescence intensity follows the triplet-triplet annihilation rate. Including the prompt contribution, the fluorescence decay may therefore be described by:
\begin{equation}
    I(t)=A_{\mathrm{P}}\exp\left(-\frac{t}{\tau}\right)
    +A_{\mathrm{DF}}
    \left(1+\frac{t}{t_{\mathrm{TTA}}}\right)^{-2},
    \qquad
    t_{\mathrm{TTA}}=\frac{1}{b[T_1]_0} \; .
\end{equation}
The amplitudes $A_{\mathrm{P}}$ and $A_{\mathrm{DF}}$ include the relevant population, fluorescence-yield, and detection factors. The delayed contribution behaves as $t^{-2}$ for $t\gg t_{\mathrm{TTA}}$, provided that other triplet-loss
processes remain negligible. Therefore, this model gives predictions surprisingly similar to the quantum mechanical approach. However, the mechanism described here is typically associated with time-independent spectral branching ratios for a singlet state relaxation. In such a case, the signals in different detection bands differ only by constant prefactors, and no spectral dependence in the general shape of the fluorescence decay curve is expected. This prediction is in clear contradiction to the experimental observations from \cite{RID_paper} (e.g., for erythrosine B -- channel 1: $\beta = 1.559 \pm 0.014$; channel 2: $\beta = 2.019 \pm 0.017$ where $\beta$ denotes the power-law exponent). Even though certain individual decay curves could be successfully described in terms of second- or even third-order processes involving triplet state annihilation, any discrepancies between individual detection channels disqualify this approach. However, importantly, the same conclusion cannot be drawn in the case of a single detection channel. In such situations, for certain power-law exponents, both quantum and delayed-fluorescence models are essentially indistinguishable. The fact that the delayed fluorescence model predicts the same temporal dependence in both bands is crucial to this reasoning. Yet, more general kinetic models involving distinct emitting populations or time-dependent spectral contributions are not excluded by this argument. Finally, the delayed fluorescence model considered here cannot explain the appearance of the power-law exponents that are greater than 2 (the particular mechanism of second- or higher-order triplet dynamics predicts only certain discrete values ranging from 1 to 2). This restriction applies to the specified kinetic model rather than to delayed fluorescence in general. Moreover, in the experiments described in \cite{Rothe:2006rma}, some substances could be described by algebraic exponents equal to 3.5 or even greater. 

\subsection{Energy-migration processes}

Other processes resulting in deviations from the exponential decay law are related to energy migration. In simple terms, an excited fluorophore can transfer its energy without the emission of a photon. This energy can then be used to excite other states. The introduction of an additional relaxation pathway leads to a corresponding quantity, the hopping rate $\nu(t)$, which affects the dynamics of the system, in particular its natural lifetime $\tau _0$. In this case, the rate equation takes the following form:
\begin{equation}
    \frac{\mathrm{d}I(t)}{\mathrm{d}t}
    = -c\nu(t)I(t)-\frac{1}{\tau_0}I(t) \; .
\end{equation}
In the model for time-dependent energy-migration processes, governed by $\nu(t)$, adopted by Włodarczyk and Kierdaszuk (for details, see Ref.~\cite{wlodarczyk2004}), deviations from monoexponential decay are described by
\begin{equation}
    \label{eq:wlodarczyk_model}
    I(t) = I_0
    \exp\left(-\frac{t}{\tau_1}\right)
    \left(1+\frac{1}{\gamma}\frac{t}{\tau_2}\right)^{-\gamma} \; ,
\end{equation}
where $\gamma>0$, while $\tau_1$ corresponds exactly to the lifetime $\tau_0$, making the notation more convenient. This model possesses particularly interesting features. Assuming that the fluorescence decay function can locally be approximated by an exponential, the effective lifetime changes from
$\tau_{\mathrm{eff}}=\tau_1\tau_2/(\tau_1+\tau_2)<\tau_1$
at short times ($t\to0$) to $\tau_{\mathrm{eff}}=\tau_1$ at long times ($t\to\infty$). Thus, the effective decay rate decreases with time, which may produce power-law-like behaviour over an intermediate time range.

\begin{figure}[!htbp]
    \centering
    \begin{subfigure}{0.48\textwidth}
        \captionsetup{width=0.89\textwidth}
        \includegraphics[width=1\linewidth]{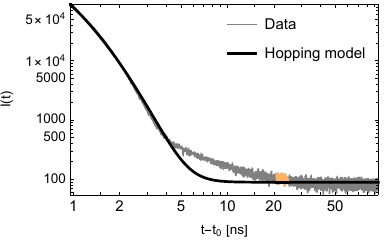}
        \caption{Erythrosine B, channel 1 -- $\chi_{\nu}^2 = 6.79$, $\gamma = 7.60 \pm 0.07$, $\tau_2 = 0.262 \pm 0.002 \, \mathrm{ns}$, $\tau_1 \to \infty$.}
        \label{fig:Wlodarczyk-erythrosine}
    \end{subfigure}
    \begin{subfigure}{0.48\textwidth}
        \captionsetup{width=0.88\textwidth}
        \includegraphics[width=1\linewidth]{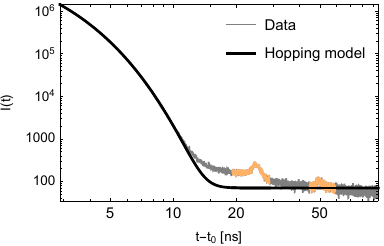}
        \caption{Eosine Y, channel 1 -- $\chi_{\nu}^2 = 9.80$, $\gamma = 73.9 \pm 0.9$, $\tau_2 = 1.0213 \pm 0.0005 \, \mathrm{ns}$, $\tau_1 \to \infty$.}
        \label{fig:Wlodarczyk-eosine}
    \end{subfigure}
    \caption{Results of the fits using the energy-migration model. Non-fitted data are marked in orange -- this is due to effects related to time-to-digital converter and its internal reference clock leading to periodic structures with the period of 25 ns.}
    \label{fig:Wlodarczyk}
\end{figure}
\medskip

For the model described by Eq.~(\ref{eq:wlodarczyk_model}), the best fit is obtained in the limit $\tau_1\to\infty$. This removes the exponential factor from the model and therefore strongly disfavours its interpretation in terms of the assumed energy-migration mechanism. Moreover, in this limit Eq.~(\ref{eq:wlodarczyk_model}) becomes
\begin{equation}
    I(t)=I_0
    \left(1+\frac{1}{\gamma}\frac{t}{\tau_2}\right)^{-\gamma} \; ,
\end{equation}
which is exactly equivalent to the Tsallis $q$-exponential of Eq.~(\ref{eq:Tsallis_fluorescence-intensity}) upon identifying
\begin{equation}
    q=1+\frac{1}{\gamma},
    \qquad
    \tau_q=\tau_2 \; .
\end{equation}
This explains why the fitted values of $\tau_2$, together with their uncertainties, are essentially identical to the corresponding values of $\tau_q$ obtained in the previous section.

\section{Conclusions}

The fluorescence decay data for both erythrosine B and eosine Y can be well described by quantum mechanical models leading to band-dependent power-law behaviour at late times (for details, see Ref.~\cite{RID_paper}). However, this interpretation is not the only possible one, as other mechanisms may also lead to power-law or quasi-power-law behaviour. The analysis presented in this work shows that the specific statistical and energy-migration parametrizations fitted here cannot reproduce the relatively sharp transition from the exponential to the power-law regime observed in the experimental data. Second-order delayed-fluorescence processes can, in principle, generate such a crossover. However, the simple delayed-fluorescence model considered here, based on a single emitting population with a time-independent emission spectrum, does not account for the observed band dependence. Our results provide further support for the interpretation of the observed late-time deviations as a~manifestation of nonexponential quantum decay. Further measurements are planned to test this interpretation.

\section*{Acknowledgements}

This work was supported by the Polish Minister of Science under the ‘Regional Excellence Initiative’ program (project RID/SP/0015/2024/01, sub-projects RID/2024/LIDER/08 and RID/2025/LIDER/02).

\bibliographystyle{unsrt}
\bibliography{bibliography}

\end{document}